\RequirePackage[2020-02-02]{latexrelease}
\documentclass[journal]{IEEEtran}

\usepackage{graphicx} 
\usepackage{amsmath} 
\usepackage{amssymb} 
\usepackage{amsfonts} 
\usepackage{algorithm} 
\usepackage{algpseudocode} 
\usepackage[utf8]{inputenc} 
\usepackage{url} 
\usepackage{hyperref} 

\makeatletter
 \DeclareRobustCommand\ref{%
    \@ifstar\@refstar\T@ref
  }%
  \DeclareRobustCommand\pageref{%
    \@ifstar\@pagerefstar\T@pageref
  }%
 \makeatother

\graphicspath{ {./figs/} } 
\begin{document}

\title{Rodent Breathing Waveforms in ApoE Rats: Statistical and Entropic Differentiation}
\author{Stephen~E.~Wormald,
        Nicholas~J.~Napoli,
        Gordon~S.~Mitchell,
        and~Alexandria~B.~Marciante
\thanks{S.E. Wormald and N.J. Napoli, are with the Department of Electrical and Computer Engineering, University of Florida, Gainesville, FL 32611 USA (e-mail: stephen.wormald@ufl.edu; n.napoli@ufl.edu).}

\thanks{A.B. Marciante and G.S. Mitchell are with the Breathing Research and Therapeutics (BREATHE) Center, College of Public Health and Health Professions, University of Florida, Gainesville, FL 32611 USA (e-mail: amarciante@phhp.ufl.edu; gsmitche@phhp.ufl.edu).}
}

\maketitle

\begin{abstract}
Apolipoprotein E (ApoE) gene variations are involved in lipid metabolism and cholesterol transport, with the ApoE4 allele being a known risk factor associated with neurodegenerative conditions later in life. Emerging evidence suggests these genetic variations may also influence respiratory function and vitality. However, the specific impact of different ApoE genotypes on breathing patterns remains largely unexplored. This work investigates differences in breathing waveform characteristics and entropy statistics derived from plethysmography (PLETH) data between rat models possessing two distinct ApoE genotypes (referred to herein as gene59 and gene95). Findings reveal significant distributional differences in common plethysmography metrics and approximate entropy between the two genotypes, observed during both active and resting states. Additionally, the study examines the transient impact of sighs (deep breaths) on these breathing metrics, demonstrating that entropy and other measures are altered in the breaths immediately following a sigh.
\end{abstract}

\IEEEpeerreviewmaketitle


\section{Background}

\IEEEPARstart{A}{ll} procedures were conducted in accordance with protocols approved by the University of Florida Institutional Animal Care and Use Committee. Experiments were performed on young adult (3-4 moths) male  humanized knock-in APOE3 and APOE4 Sprague-Dawley rats (Envigo; IN, USA). Rats were housed in pairs at 24$^{\circ}$ C with a 12/12 light/dark cycle (lights on: 06:00; lights off: 18:00) with access to food and water ad libitum. All rats underwent a 14-day acclimation period prior to experiments. 

The Apolipoprotein E (ApoE) gene plays a critical role in lipid metabolism and the transport of cholesterol throughout the body \cite{mahley1988apolipoprotein}. Different alleles of this gene exist, with the ApoE4 variant being significantly associated with an increased risk for several neurodegenerative diseases, most notably late-onset Alzheimer's disease, in human populations \cite{eh1993gene}. While the link between ApoE4 and neurodegeneration is well-studied, its potential influence on other physiological systems, including respiratory control, is less understood. There is preliminary evidence suggesting that the ApoE genotype might impact breathing, potentially affecting neural control elements such as chemoreflexes or the overall stability of breathing patterns \cite{gottlieb2004apoe, kadotani2001association}. Such instability is sometimes characteristic of breathing patterns observed in humans and rodent models of neurodegenerative diseases \cite{varga2016reduced,marciante2023progressive}. Given these associations, understanding how different ApoE genotypes relate to baseline breathing characteristics is crucial.

This study analyzes a Plethysmography (PLETH) dataset from rodent models to explore differences in breath characteristics associated with genotype and activity state. The dataset comprises recordings from twenty rats, allowing for comparisons based on activity state (active vs. resting periods; Comparison 1) and ApoE genotype (referred to as gene59 vs. gene95 for blinding purposes; Comparison 2).

\begin{figure}
    \centering
    \includegraphics[width=0.5\textwidth, trim=8cm 7cm 7cm 6.5cm]{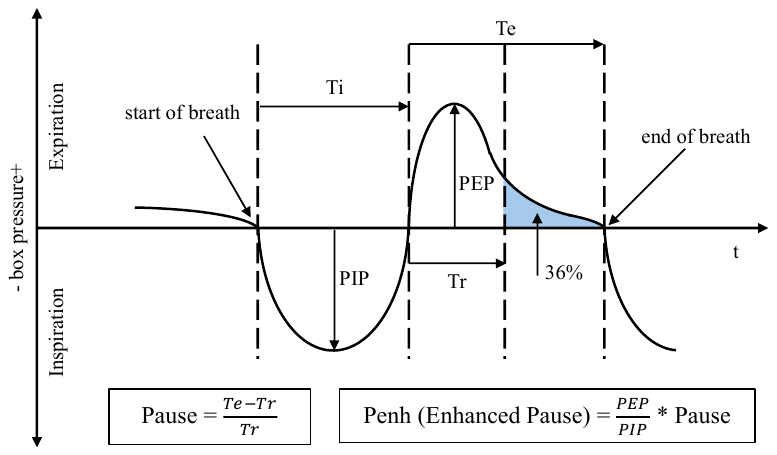}
    \caption{PLETH breath measures, as determined and replicated from \cite{Bates1985}. The primary measures include the inspiration duration (Ti), expiration duration (Te), time to expire (Tr), the Peak Inspiratory Pressure (PIP), Peak Expiratory Pressure (PEP), breath Pause (Pause), and enhanced pause (Penh).}
    \label{fig:plethstats}
\end{figure}

The primary research question is therefore to identify what differences exist in the statistics of breaths extracted from the PLETH data for the two comparison categories. This question is divided into three null hypothesis: (1) There are no difference in the distribution of plethysmography breath metrics for Comparison 1 and Comparison 2. Common metrics include inhalation time (Ti), expiration time (Te), and others defined by Bates \cite{Bates1985} (Figure \ref{fig:plethstats}). (2) There is no difference in the distribution of waveform complexity, measured using approximate entropy, for Comparison 1 and Comparison 2. (3) There is no difference between the waveform statistics and entropy measures before and after a sigh waveform, or a breath where the rodent inhales and exhales deeply, for both Comparison 1 and Comparison 2. The following sections overview the methodology to test these hypothesis and present relevant findings and conclusions.

\section{Methods}
\label{sec:methods} 

The PLETH data was collected using the IOX2 software using a sampling frequency of 1000hz. The raw PLETH waveform was exported from the IOX2 software using a software extension for saving EDF files. All analysis was performed in the python programming language. 

In this analysis, the rat genotype was blinded to remove bias in analysis, and are referred to as the gene59 and gene95 variations. However, the rat activity type (active and resting) was known, where midactive refers to the rats in their awake cycle (12 AM), and midrest (12 PM) refers to rats in their sleep cycle. These categories produce four rodent scenario types: (1) Midactive-59, containing six experiments, (2) Midactive-95, containing two experiments, (3) Midrest-59, containing six experiments, and (4) Midrest-95, containing six experiments. There are fewer runs in the Midactive-95 category as there was a recording error during the experiments.

The methodology revolves around three main stages. The starting task is to extract the breath waveforms from the PLETH data. This work prefaces two subsequent tasks: (Task 1) Comparing the breath and entropy statistics between the genetic and activity categories (per hypothesis 1 and 2. (Task 2) Compare the breath and entropy statistics before and after the rats take a deep sigh, per hypothesis 3. The full process is visualized in the flowchart in Figure \ref{fig:approach}. The methodology pertaining to each task is explained in the following subsections: 
\begin{figure*}[h!]
    \centering
    \includegraphics[width=15cm]{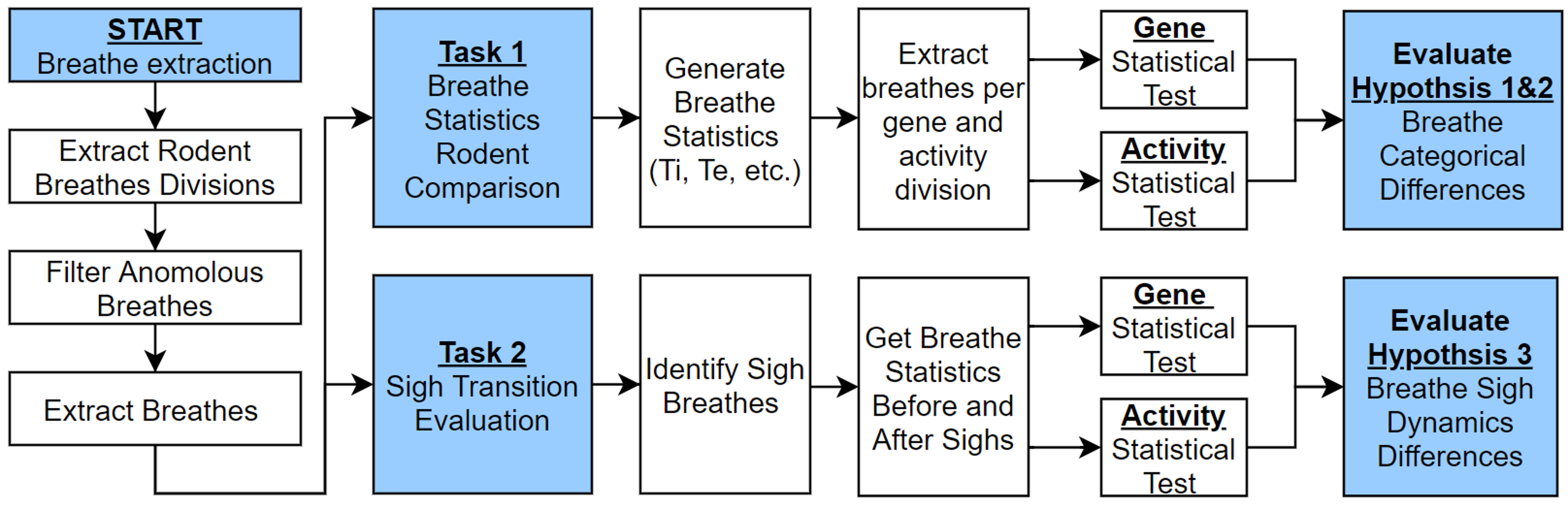}
    \caption{Methodology flowchart, breaking the main lines of work into two main task lines focused on different hypotheses.}
    \label{fig:approach}
\end{figure*}

\subsection{Rat Breath Extraction and Filtering}
\label{sec:breathmethod} 

An example of the PLETH data is shown in Figure \ref{fig:preprocess}A, where Figure \ref{fig:preprocess}B shows an example of different features in the raw PLETH data. Note there are nomial breaths surrounded by larger rodent sighs, high frequency waves believed to be associated with rat sniffing behavior, and then salt and pepper noise. The goal is to remove noise before segmenting the individual breaths. The breath segmentation contained several steps: (1) salt-and-pepper noise filtering, (2) zero-crossing flagging, (3) waveform anomaly detection, and (4) breath database formation.  

\subsubsection*{Salt-and-Pepper Noise Filtering}
\begin{figure*}[h!]
    \centering
    \includegraphics[width=15cm]{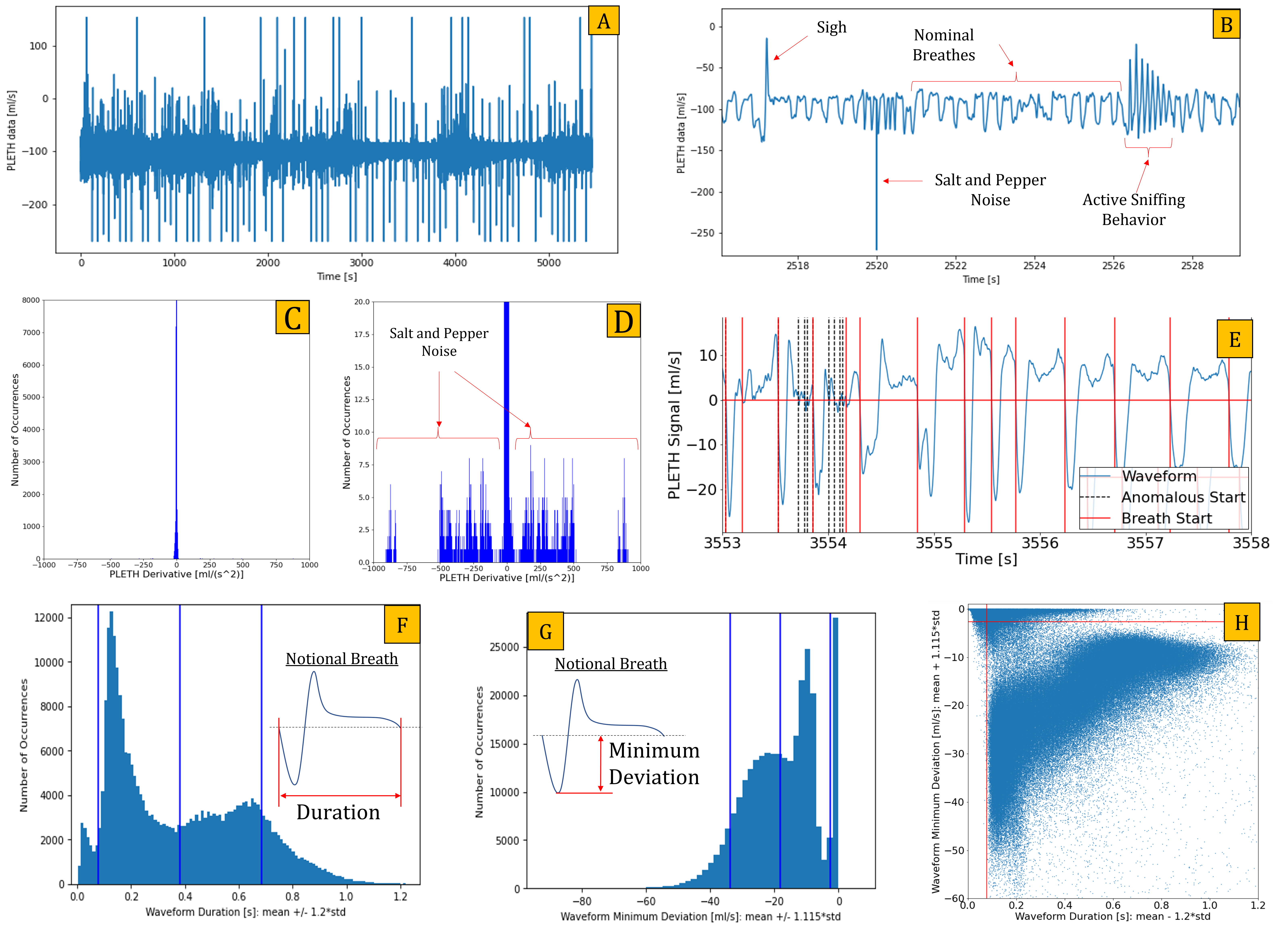}
    \caption{Example of PLETH data breath pre-processing. (A) The full duration of data from one data collection run. (B) A zoomed in region of the full experiment, showing standard breathing, a sigh breath, and rodent sniffing behavior as the rodent sometimes gets closer to the pressure transducer, and salt-and-pepper noise in the raw signal. (C,D) Distribution of the signal derivatives, zoomed out and zoomed in respectively. Note for the 109239980 samples, the mean was 1.84e-06 ml/s, the standard deviation was 1.81 ml/s, and the minimum and maximum values are -907 and 907 ml/s, respectively. (E) Start times of each waveform shown as vertical red lines, where nominal and anomalous divisions were determined by the thresholds in the breath duration distribution in (F), and the breath minimum deviation distributions in (G). The thresholds used are visualized in (H) to remove breaths divisions associated with the metric cluster in the upper left hand corner.}
    \label{fig:preprocess}
\end{figure*}

The salt-and-pepper filtering was implemented as a conditional moving average filter. While a moving average filter acts over all points in a time-series signal, the conditional approach identifies locations where the signal derivative is anomalous, and only then replaces the anomaly with an average of the points surrounding the anomaly. This approach preserves the original signal where feasible, but has a two main downsides: (1) a hyperparameter is added that needs to be tuned ($dP_{threshold}$, which changes the prevalence of flagged anomalies), and (2) it is possible to miss salt-and-pepper noise if the hyperparameter is ill-defined. Via visual inspection, a value of $dP_{threshold} = 9$ removed all significant salt-and-pepper noise, though later smaller salt-and-pepper noise was observed in the signal. A median filter could be used to resolve this issue in future work. The conditional moving algorithm is detailed in Algorithm \ref{algorithm:sapalg}. Figure \ref{fig:preprocess}C and Figure \ref{fig:preprocess}D show the distribution of the salt and pepper noise. Note the mean is near zero, and the standard deviation (see figure caption) is small, which seems to indicate that derivative outliers past the +/-100 $ml/s^2$ regions are far outside the distribution of a normal derivative value. These outliers were removed according to Algorithm \ref{algorithm:sapalg}.

\begin{algorithm}
\caption{Salt and Pepper Filter}\label{algorithm:sapalg}
\begin{algorithmic}[1]
    \State Read Rat Plethysmography Signal: $s $
    \State Set Filter Threshold: $dS_{threshold} = 9 $
    \For{$i = 1$ to $length(s)-1$}
        \State $dS_{i} = s_{i}-s_{i-1} $
    \EndFor    
    \State $dS_{mean} = \frac{\sum{dS_{i}}}{length(dS)} $
    \State $dS_{std} = \sqrt{\frac{\sum{(dS_{i} - dS_{mean})^2}}{length(dS)}} $
    \For{$ i = 2$ to $length(s)-2$} \Comment{Truncates s}
            \If{$dS_{i+1} >= dS_{mean} + dS_{threshold}*dS_{std}$}
                \State $s_{i} = (s_{i-2} + s_{i+2})/2 $
            \EndIf
    \EndFor    
    \State \textbf{return} $s$
\end{algorithmic}

\end{algorithm}

\subsubsection*{Zero-Crossing Flagging} 
After removing the salt and pepper noise, each PLETH data stream was centered about the mean signal ($s_{centered} = s_{rodent} - s_{mean}$). The start of inhalation and expiration were flagged when the PLETH became negative and positive, respectively. This approach follows work by Micheal in Reference \cite{Michael2021}. However, as seen in Figure \ref{fig:preprocess}E, there are some breath divisions which seem to split breaths in the middle of the expiration period due to noise in the signal. These anomalies were detected and removed.   

\subsubsection*{Breath Anomaly Detection} 
To remove anomalies in the breath divisions, two metrics were calculated: (1) the breath duration, and (2) the breath minimum deviation. The distribution of these metrics are plotted in Figure \ref{fig:preprocess}F and Figure \ref{fig:preprocess}G. Each distribution is non-Gaussian, and displays several peaks. For each metric, there is a small peak near the zero-value where there is a higher density of anomalous breathes. Figure \ref{fig:preprocess}H shows these metrics plotted against each other, revealing two primary breath clusters. The red lines in this figure show two univariate thresholds which were used to remove anomalous breath divisions. Figure \ref{fig:preprocess}E shows black dotted lines where anomalous breath divisions were removed. Ideally, the secondary cluster of metrics in Figure \ref{fig:preprocess}H would be verified as being anomalous using metrics published in literature. Furthermore, the threshold could be improved by using either a multivariate linear decision boundary or a non-linear clustering based approach. These approaches were explored but not used in the final analysis to aid reproducibility in the method. 

\subsubsection*{Rat Breath Database Formation}
After removing outlier breaths from the dataset, the start and end times were used to find the plethysmography data associated  with each breath. The rodent number, activity type, gene type, breath start time, breath end time, and raw signal values were saved to a database for the statistical characterization performed in Section \ref{sec:statmethod}. Note the signals were down-sampled by a factor of ten, from a sampling frequency of 1000hz to 100hz, to save space and reduce processing time. All rat breaths were zero-padded to a length of 400 samples. 

\subsection{Statistics Calculation and Comparison}
\label{sec:statmethod} 

Two main categories of statistics were calculated for all rat breaths: (1) the plethysmography statistics in Figure \ref{fig:plethstats}, and (2) the approximate entropy. These statistics are summarized below: 

\subsubsection*{Plethysmography Statistics} The metrics in Figure \ref{fig:plethstats} were calculated as follow for each rat breath, where $s$ is the PLETH signal: 

\begin{equation}
\begin{split}
Ti = Ti_{end} - Ti_{start}                \\ 
Ti_{end} = End\:time\:of\:inspiration     \\
Ti_{start} = Start\:time\:of\:inspiration \\
\end{split}
\label{equation:Ti}
\end{equation}

\begin{equation}
\begin{split}
Te = Te_{end} - Te_{start}               \\
Te_{end} = End \:time\:of\:expiration    \\
Te_{start} = Start\:time\:of\:expiration \\
\end{split}
\label{equation:Te}
\end{equation}

\begin{equation}
\begin{split}
Tr = Tr_{end} - Te_{start}               \\
Tr_{end} = Time\:of\:36\%\:remaining\:expiration\\    
\end{split}
\label{equation:Tr}
\end{equation}

\begin{equation}
\begin{split}
PIP = minimum(s_{i}) \\
Ti_{start}>=i>=Ti_{end}\\ 
\end{split}
\label{equation:PIP}
\end{equation}

\begin{equation}
\begin{split}
PEP = maximum(s_{i}); \\
Te_{start}>=i>=Te_{end} \\
\end{split}
\label{equation:PEP}
\end{equation}

\begin{equation}
Pause = \frac{Te - Tr}{Tr}
\label{equation:Pause}
\end{equation}

\begin{equation}
Penh = \frac{PEP}{PIP}*Pause
\label{equation:Penh}
\end{equation}  

\subsubsection*{Entropy Statistics} The Approximate Entropy (AE) was calculated to measure the complexity (or signal randomness) of rat breaths. AE was the only entropy measure calculated, though more entropy methods could be tested in subsequent work. An implementation of AE from the EntropyHub python library was used for this work \cite{approximate_entropy}, which takes two inputs: the Radius Distance Threshold ($r$), and the Embedding Dimension ($m$). The embedding dimension was changed from 0-4 in this work to characterize the complexity over different template lengths, and $r$ was set at the default value of $0.2*std(s)$. This results in 5 entropy measures, one per embedding dimension. 

The plethysmography and entropy statistics were calculated for each breath, and were used to evaluate Comparisons 1 and 2 for each hypothesis. As seen in the results section, the statistics distributions were bimodal, making the two-sample t-test unfit for statistical comparison. The  two-sample Kolmogorov-Smirnov test was selected as this approach does not require compared distributions to be Gaussian. The distributions and results from statistical comparisons are shown in the results section.

\subsection{Sigh-Impact on Breath Statistics}
\label{sec:sighmethod} 

\begin{figure*}[htbp]
    \centering
    \includegraphics[width=15cm]{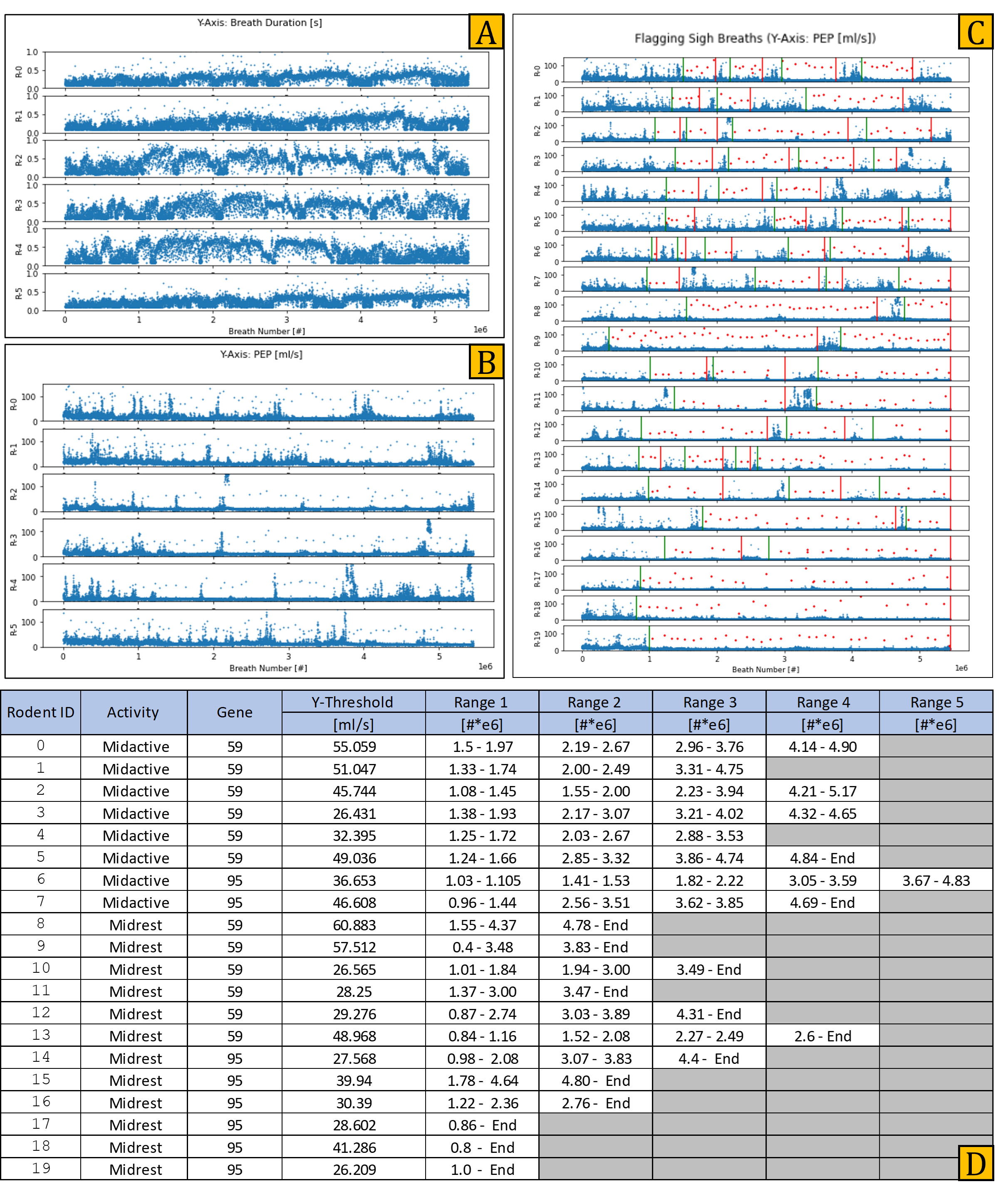}
    \caption{(A) Breath duration plotted with respect to the breathe number for rats 1-5, showing how there are periods where the breath duration is longer corresponding to rat inactivity. (B) The PEP over time, showing PEP outliers where sighs occurred. (C) Rat sighs flagged from the breath sequence, but from the rodent resting states, as defined manually and summarized in (D) along with the PEP thresholds per rat.}
    \label{fig:sigh_extract}
\end{figure*}

 This section considers how a rat sigh may impact breathing statistics after the sigh occurs. The methodology examines whether the plethysmography and entropy statistics change between the pre- and post-sigh scenarios. Key steps in this methodology include: (1) differentiate the sigh breaths from nominal breaths, (2) identify the breaths and corresponding statistics occurring N-steps before and after each sigh, and (3), perform a statistical test on the pre- and post-sigh statistics to evaluate differences in both the Comparison 1 and Comparison 2 categories. These steps are detailed below: 

 Figure \ref{fig:sigh_extract}A shows the breath-duration statistic over with respect to the breath number. Notice there are periods in the rodent behavior where the duration increases for a period. These regions are associated with states of rodent inactivity, as rat breaths become longer when the rodent is calm, not exerting much energy, and therefore requiring less frequent breathing. Figure \ref{fig:sigh_extract}B shows the PEP metric of each breath, where outliers appear with a greater magnitude, and are generally more visible when the breath duration increases and the rodent is resting (seen in Figure \ref{fig:sigh_extract}A). The PEP outliers are sigh breaths. These breaths were extracted by filtering the breaths based on the PEP value in the rodent rest periods. The time of each rest and the PEP threshold used for each rodent is summarized in Figure \ref{fig:sigh_extract}D. Some rodents were more active than others, making it beneficial to include target specific regions using additional breath ranges. Figure \ref{fig:sigh_extract}C highlights the flagged sigh breaths in red. The green and red lines indicate the start and stop of a rest period, respectively.

\begin{figure*}[ht]
    \centering
    \includegraphics[width=15cm]{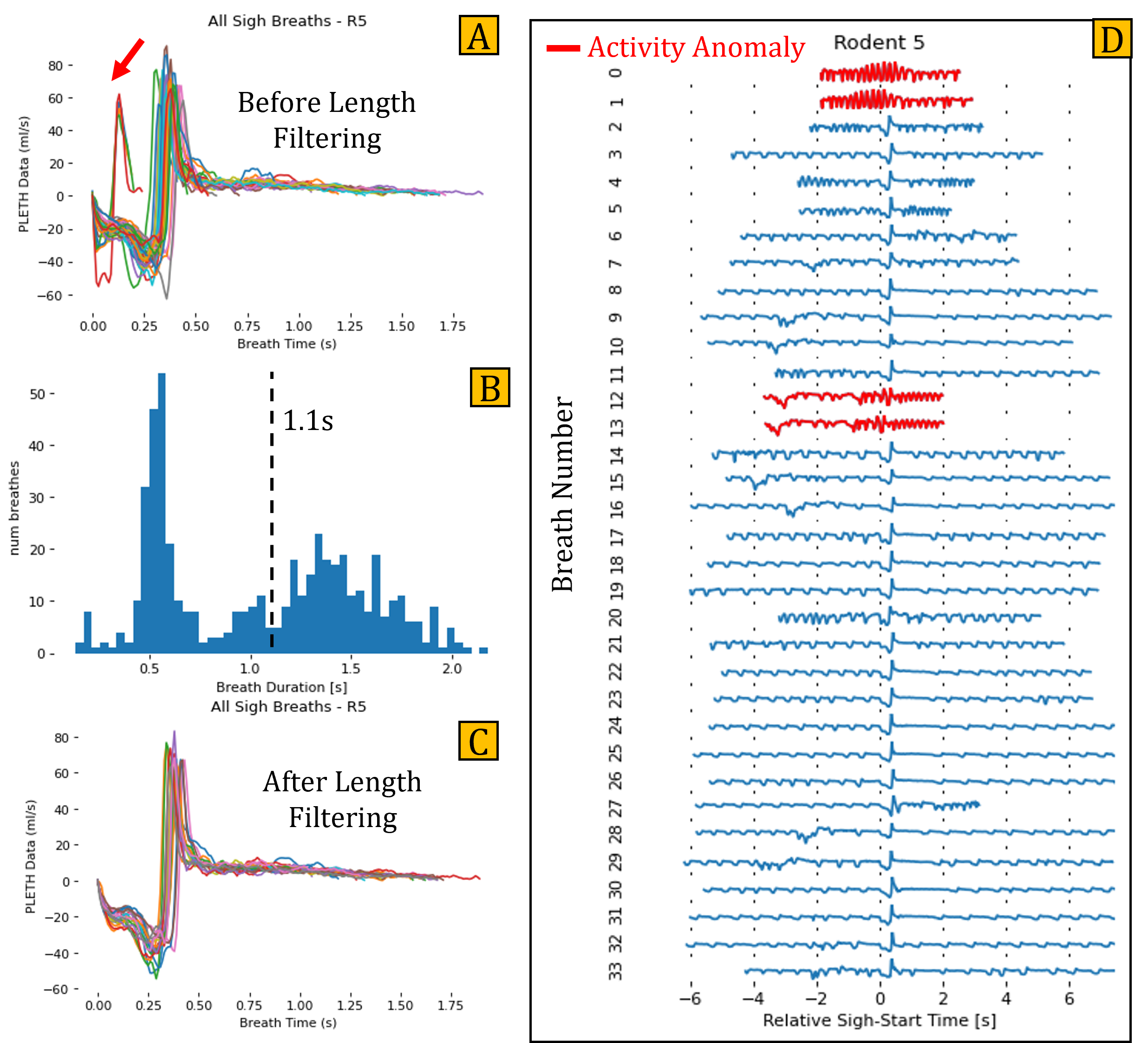}
    \caption{(A) Visualization of sigh breaths extracted from the full breath sequence. Notice how there are some waveforms which do not appear to be sighs, but are shorter in duration. (B) Distribution of breath duration of the flagged sighs. A threshold of 1.1 seconds was selected to filter out breaths that were likely not sighs. (C) Resulting sigh breaths after duration filtering. (D) Visualization of 10 breaths before and after the sigh breaths where several waveforms were flagged manually as being anomalous and removed from subsequent analysis. This process was manually performed for all rats.}
    \label{fig:sigh_sequence}
\end{figure*}

 Figure \ref{fig:sigh_sequence}A shows the sigh breaths extracted for rat 5, which is a midactive rat with the gene59 variation. Note there are short breaths in this figure. These short breaths are often not breaths, though they crossed the PEP threshold. Therefore, the sighs were filtered based on their duration. The distribution of breath duration is shown in Figure \ref{fig:sigh_sequence}B, where the threshold used (1.1 seconds) is depicted. Figure \ref{fig:sigh_sequence}C shows the filtered breaths after duration filtering. Figure \ref{fig:sigh_sequence}D plots the sigh breaths with 10 waveforms before and after the sigh. Note the sigh is placed in the middle of the figure, showing a large spike where the rodent is exhaling (the PEP location). Despite the duration filtering, several anomalous patterns were observed that did not seem to represent sigh periods. These anomalies were flagged in red and removed manually for all 20 rats. Having obtained the sigh period for each rat, pre- and post-sigh comparisons were then pursued. 

 Hypothesis 3 seeks to understand how breath statistics change after a sigh occurs. Two methods were used to evaluate how statistics changed along the duration of a sigh period. The first used the breath statistics from all breaths $M$ before or after the sign, and plotted them as a box and whisker plot. The result shows how a breath statistic changes across the sigh sequence, and focuses on trends observed across all rats. The second method is tabular, and focuses on understanding how pre- and post-sigh metrics differ in the Comparison 1 and Comparison 2 categories. For each comparison type and the breath metrics from all breaths pre- or post-sigh were obtained. In each scenario, the distributions were compared using a two-sample t-test as the distributions were nearly Gaussian, resulting in a p-value for both the pre- and post-sigh situations. These results show how the comparison types differ at different stages of the sigh sequence. The difference between the pre- and post-sigh p-values was also calculated to show how the to evaluate how the breath comparisons change over time.

\section{Results}
\label{sec:results}

The results from this work are organized according to the three hypotheses discussed in the introduction: 

\subsection{Hypothesis 1: Breath Statistics}
\begin{figure*}[h!]
    \centering
    \includegraphics[width=15cm]{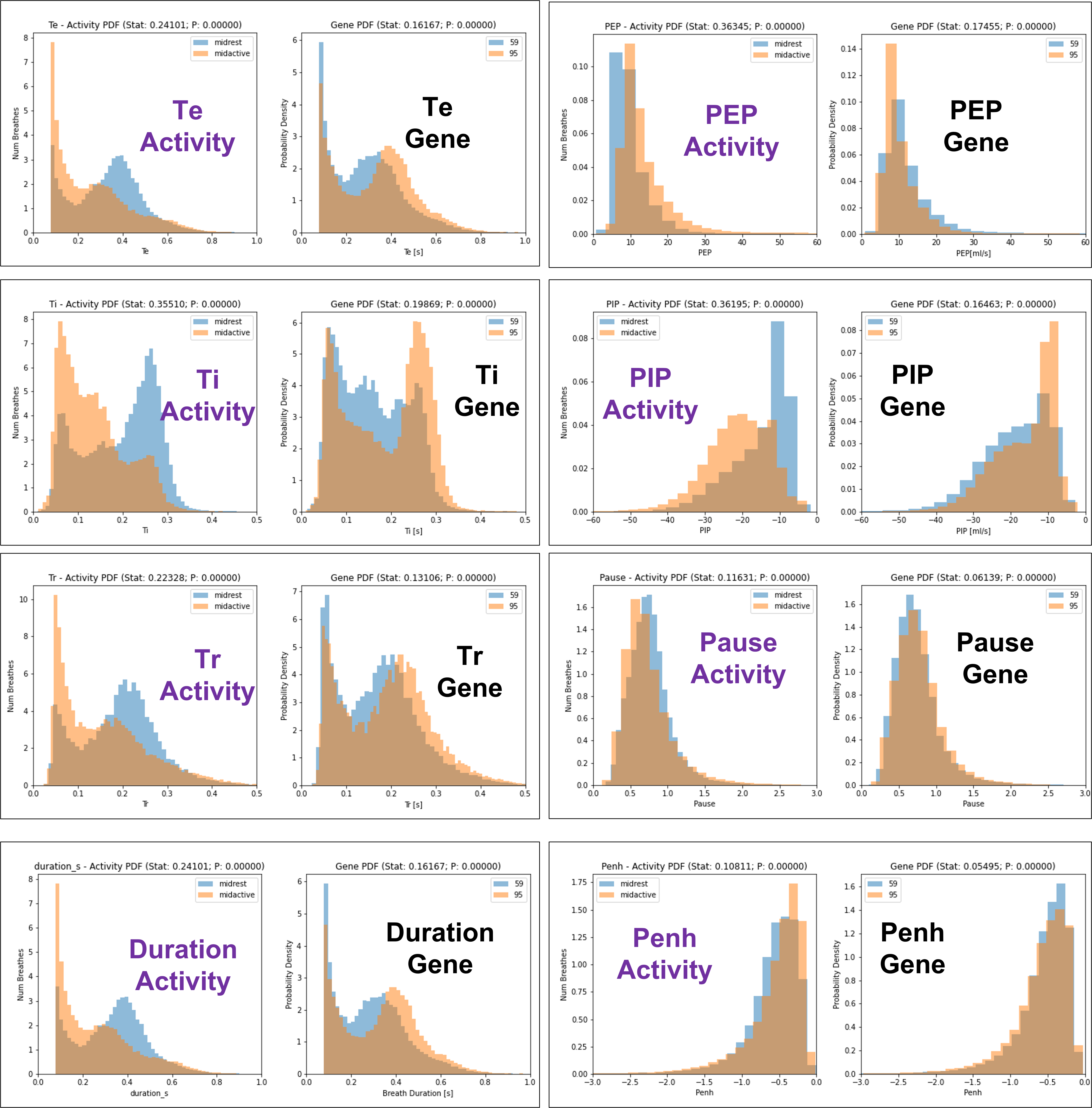}
    \caption{Distributions of plethysmography statistics for the Comparison 1 and Comparison 2 types. Each figure is labeled according to the statistic and comparison type name.}
    \label{fig:breath_res}
\end{figure*}

 Figure \ref{fig:breath_res} shows the distributions of each breath statistic used to perform Comparisons 1 and 2 (activity comparison and genetic comparison, respectively). Each subplot is labeled with the statistic name and the comparison type. The purple labels indicate the activity comparison (active and resting states), and the black labels indicate a genetic comparison (gene95 and gene59). All statistical tests returned a P value of 0.00000 (up to 5 decimal places), which was at first surprising, but was verified which by testing the function using different distributions. These results indicate there is a statistically significant difference between the means of each breath statistic for both the Comparison 1 and Comparison 2 categories.  
\label{sec:h1} 

\subsection{Hypothesis 2: Breath Entropy }

\begin{figure*}[h!]
    \centering
    \includegraphics[width=15cm]{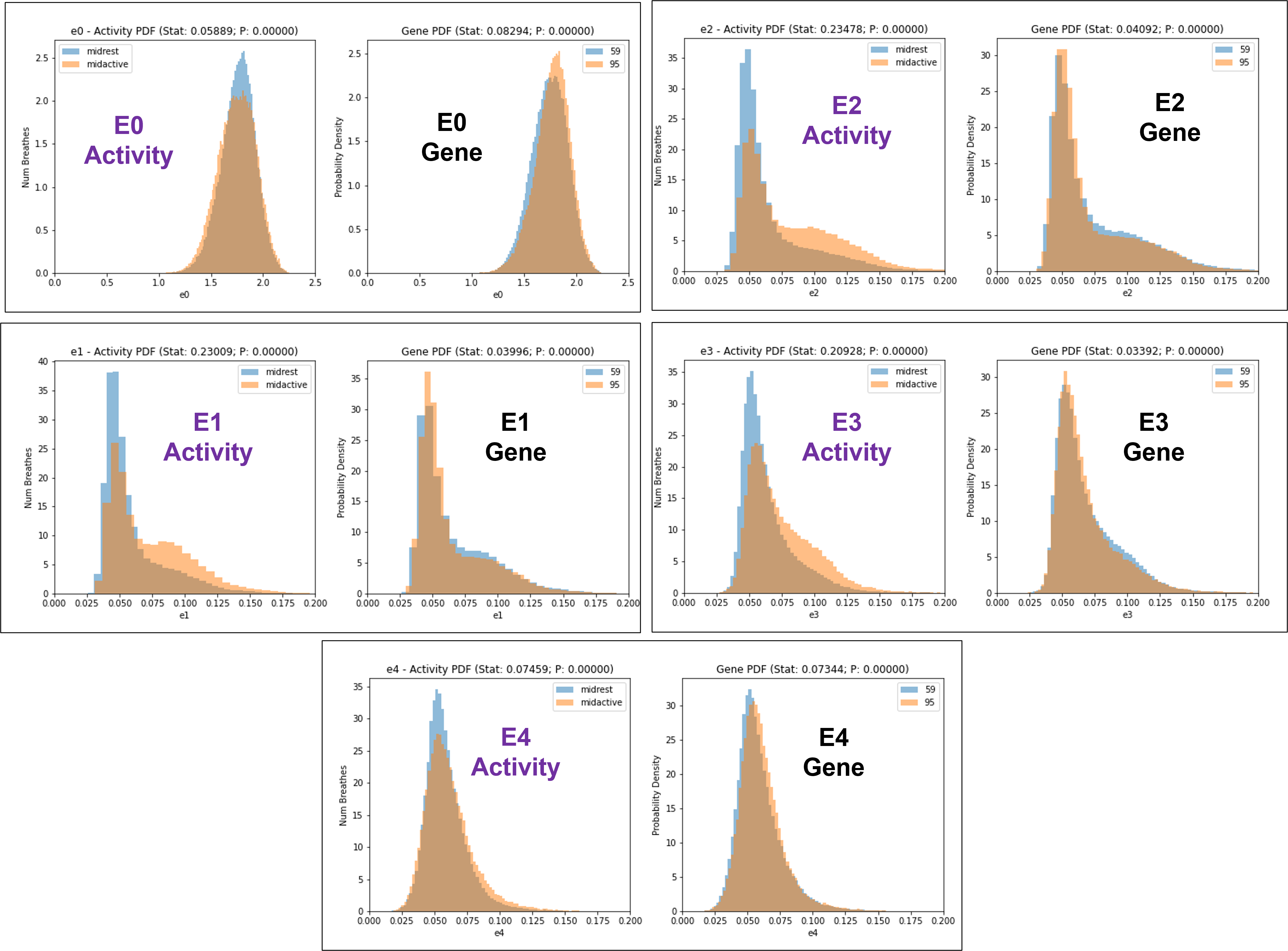}
    \caption{Entropy statistic distributions for the activity and genetic comparisons.}
    \label{fig:entropy_res}
\end{figure*}

 Figure \ref{fig:entropy_res} compares the entropy statistic distributions across the Comparison 1 and Comparison 2 types. Differences may be examined with respect to each category type. As with the plethysmography metrics, the statistical test returned a p-value of value of 0.00000 (up to 5 decimal places) for each comparison. One reason for the small p-value is the large number of breaths used in the comparison. 

\label{sec:h2} 

\subsection{Hypothesis 3: Sigh Breath Sequence }
\label{sec:h3} 

 Figure \ref{fig:sigh_res} shows how the plethysmography and entropy metrics changes for each waveform preceding and proceeding the rat sigh. Each plot shows pre-sigh breaths (breaths 1-10), the rat sigh (breath 11), and post-sigh breaths (breaths 12-21). Note each plot contains data across all rats, regardless of their genetic and activity differences, in an attempt general changes that occur before and after the sigh. Table \ref{table:sigh_pleth_res} and Table \ref{table:sigh_ent_res} look at the differences in Comparison 1 and Comparison 2 for the pre- and post-sigh breaths.  

\section{Discussion}
\label{sec:discussion}

Discussions are given with respect to each hypothesis: 

\subsection{Hypothesis 1: Breath Statistics}
 Differences between the distributions in Figure \ref{fig:breath_res} may be discussed with respect to the activity and genetic comparisons. For Comparison 1 (activity comparison), the difference between resting and active behavior can be determined by examining the general differences across each plethysmography statistic: (1) There are more long-duration breaths in the midrest category than the midactive category. This finding makes sense as rodents who are resting tend to breathe more slowly and rhythmically than active rats who are more hyperactive in their sniffing behavior. (2) The Ti, Te, and Tr metrics are longer, which makes sense as they are directly proportional to the duration of the waveform. (3) The PEP and PIP metrics are generally lower (closer to zero) for the midrest case, indicating that resting rats may be inhaling and exhaling over a longer period of time rather than sucking air into their lungs quickly. 
 The genetic comparison can be performed similarly. The the gene59 category tends to have shorter duration statistics when compared to the gene95 category, including the overall duration, Te Ti and Tr statistics. The distributional differences do not seem to be as large for the PEP, PIP, Pause, and Penh metrics. However, the peak expiration and inspiration generally seems to be larger for the gene59 condition. 

\subsection{Hypothesis 2: Breath Entropy} 
 For the activity comparison (Comparison 1), the active rodents seem to have higher entropy values across most of the entropy metrics. Specifically, the E1, E2, and E3 distributions have noticeably larger values. In theory, active rats will likely have larger variation in their breathing behavior from breath to breath as compared to rats who are resting and laying still. While the p-value is zero to 5 decimal places for each metric, it is unclear why the entropy is different for Comparison 1. For example, the entropy may be different because there is a larger variation in the types of waveforms, or it could result from increased noise in the waveform as the rodent moves around the pressure transducer. 
 The entropy distributions may be compared for the genetic cases as well (Comparison 2). While the p-values indicates statistically significant differences between the means across each entropy value, the distributions look more similar between the gene59 and gene95 cases than for the activity comparison. The p-value may be small as a result from the high number of waveforms used to build the distributions, as there were more than 100,000 breaths in each distribution.

\subsection{Hypothesis 3: Sigh Breath Sequence}
\begin{figure*}[h!]
    \centering
    \includegraphics[width=15cm]{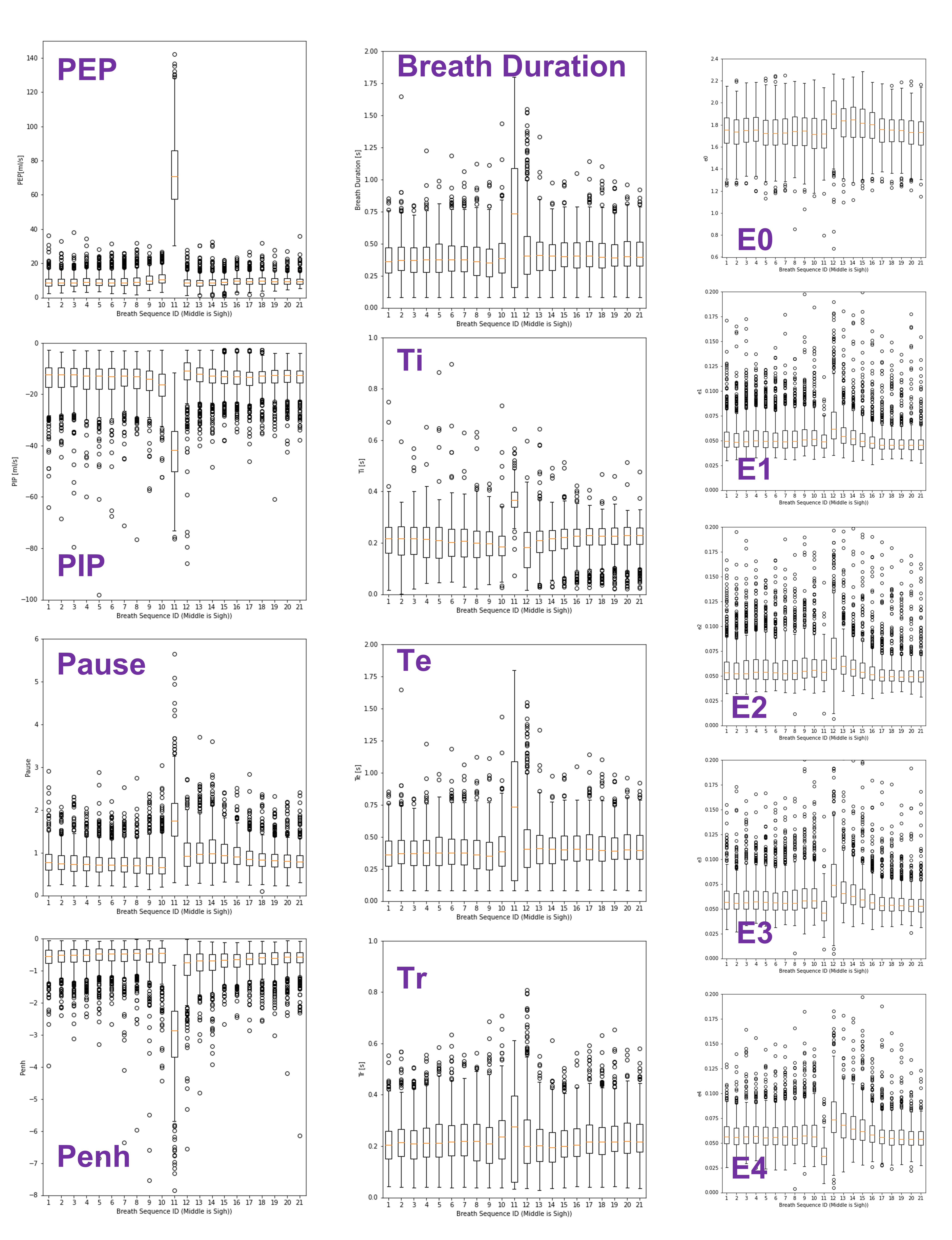}
    \caption{Breath and entropy statistic distributions plotted with respect to the breath number in the breath sigh sequence. Breaths 1-10 are pre-sigh breaths, breath 11 is the sigh, and breaths 12-21 are post-sigh breaths.}
    \label{fig:sigh_res}
\end{figure*}

 Some breath statistics change more noticeably from the pre- to post-sigh periods, while others are seemingly unchanged. For example, in Figure \ref{fig:sigh_res}, the PEP, PIP, breath duration, Ti, Te, and Tr metrics seem relatively unaffected by the sigh. However, the Pause and Penh metrics seem somewhat affected, and the entropy values are most affected. Looking at the E4 entropy distributions, there is a consistent entropy until the rat sighs, during the rodent sigh the entropy decreases, and after the sigh the entropy increases sharply before gradually decaying to a steady state value. The reduced complexity of the sigh makes sense as there is less variability through the duration of the entire waveform, making the breath relatively flat and less complex. The increased entropy after the breath may result from variability in the kind of breaths that occur immediately after a sigh, even if the breath type converges to a common type with consistent statistics. 
 Table \ref{table:sigh_pleth_res} and Table \ref{table:sigh_ent_res} can be used to evaluate Comparisons 1 and 2, and how differences between the activity and gene types change in the pre- and post-sigh regions. For example, the top four rows compare the differences in breath duration for both the genetic and activity cases. When focusing on the genetic comparison type, we can see the two categories, gene59 and gene95 shown in the ``Category 1" and ``Category 2" columns of the table, respectively. The mean and standard deviation of each of these categories is reported, along with distribution comparison metrics such as the means-difference and the P value. Using these metrics, a p-value of 0.000 is found for the pre-sigh, genetic comparison, indicating that there is a difference in the breath duration between the gene59 and gene95 distributions prior to the sigh. Note how this finding agrees with the results from Figure \ref{fig:breath_res}. Similarly, a p-value of 0.000 is also found for the post-sigh, genetic comparison. In other words, there is no significant difference between the pre-sigh and post-sigh distribution comparisons. Said differently, the sigh has little impact on the distribution comparison. However, the sigh impact is higher for other variables. 
 There is a large sigh impact between the pre- and post-sigh comparisons for the Pause metric, indicating that the midactive and midrest rats may vary in their sigh-dynamics. Examining these statistics further shows that the Pause statistic is similar between the active and resting cases before the sigh, but are dissimilar after the sigh occurs, likely because the standard deviation of the Pause statistic becomes larger for the active case, but is relatively constant for the resting case. In this way, the sigh impact metric can highlight differences in how breath metric dynamics are different between categories. However, it remains unclear why the variability of the Pause would increase for the active case but not for the resting case.
 There is also a high sigh-impact for the genetic comparison, specifically for the E1 and E2 measures. As each measure describes similar information, attention can be given to the E2 metric. In this case, the breaths’ entropy is more similar between the genetic conditions after the sigh than before. In this case, the gene59 mean entropy becomes larger across the sigh, though the gene95 mean remains the same. The reason for this difference is challenging to tell as the gene type is blinded.  

\section{Conclusion}

\label{sec:conclude}
Salient conclusions are included with respect to the three main hypotheses and key stages of analysis:

\subsection*{Hypothesis 1 - Breath Statistics:}
\begin{itemize}
    \item There is evidence to reject the null hypothesis that the breath statistics are not different between the genetic conditions, and activity conditions. There are statistically significant differences in the breath statistics between both the active and resting rats, and the gene95 and gene59 rats. 
    \item The resting rats tend to have longer breaths, inspiration, and expiration metrics as compared to the active rats. They also tend to have smaller peak inspiratory and expiratory pressures (PIP and PEP), and longer pauses between breaths, which may be related to the increased breathing duration. 
    \item The gene59 rats tend to have shorter breath cycles, inspiration, and expiration times as compared to the gene95 rats. They tend to have larger peak inspiratory and expiratory pressures (PIP and PEP), yet with slightly shorter pauses between breaths. The reason for these trends is unknown as the gene type is a blinded variable.     
\end{itemize}

\subsection*{Hypothesis 2 - Breath Entropy:}
\begin{itemize}
    \item There is evidence to reject the null hypothesis that the entropy statistics are not different between the genetic conditions, and activity conditions. There are statistically significant differences in the entropy statistics between both the active and resting state rats, and the gene95 and gene59 rats. However, the differences are more pronounced between the active and resting rats.
    \item The approximate entropy, or complexity, of breaths is higher for the active rats than for the resting rats. The exact cause is unknown, but may be related to more variable breathing patterns when active, or a higher prevalence of different waveform types when active that may be associated with sniffing or moving around.
    \item Some difference between rat breath entropy is identified between the gene95 and gene 59 rats, though the difference is less visibly pronounced than for the active and resting comparison.
\end{itemize}

\subsection*{Hypothesis 3 - Sigh Breath Sequence:}
\begin{itemize}
    \item There is some evidence to reject the null hypothesis that the breath and entropy statistics are the same before and after a rat sighs, though not for all metric, genetic type, or activity type comparisons. 
    \item The Pause, Penh, and entropy statistics change between the pre- and post-sigh breaths. However, the significance of the difference seems to depend on how many breaths from the sigh are being considered.
    \item The pre- and post-sigh statistics distributions seem to be the most different between the resting and active rodents for the Pause metric, and most different between the gene95 and gene59 rats for the E2 and E3 entropy statistics. It is unclear what influences these pre- and post-sigh differences. 
\end{itemize}

Salient next steps for this research may include:
\begin{itemize}
    \item Determine prevalence of waveform types in the active and resting, and gene95 and gene59 cases. Waveform types may be identified using clustering methods.   
    \item Study how the significance of pre- and post-sigh statistic differences change as a function of the number of breaths away from the sigh. Plot the p-value as a function of number of breaths away from the sigh. 
    \item Consider the probability of transitioning to and from different breath types over time. 
    \item Extend the wave type prevalence to the automatic identification of active and resting states. 
    \item Incorporate different entropy measures, such as permutation and multi-scalar entropy.  
    \item Change the salt-and-pepper filter to be a median filter rather than a conditional moving average.  
\end{itemize}

\section{Funding}

This work was supported by NIH K99HL175029 (ABM) and NIH R01HL149800 (GSM).

\begin{table*}[h!]
    \centering
    \includegraphics[width=15cm]{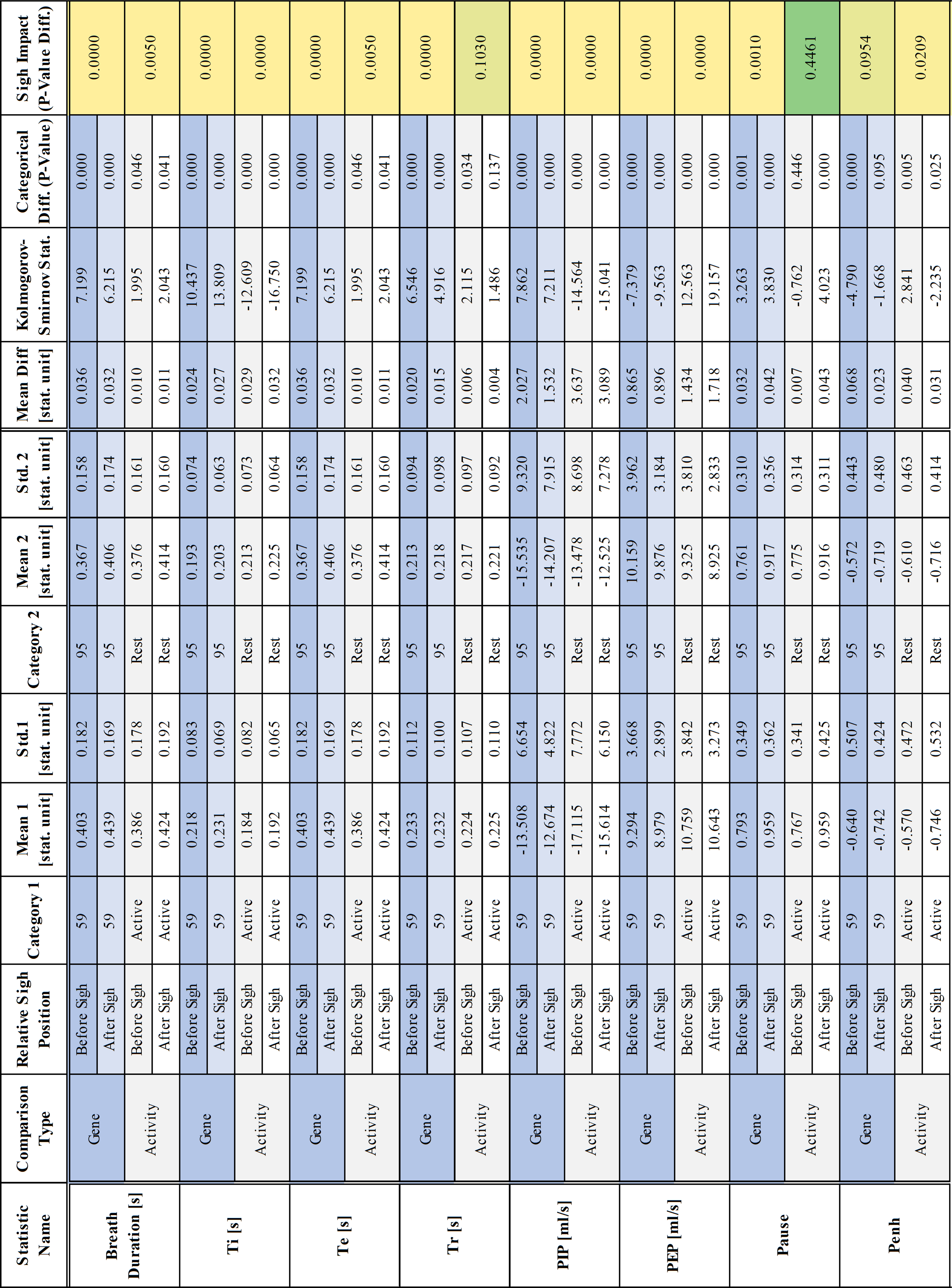}
    \caption{Comparison of the breath statistics before and after the rat sigh for both genetic and activity type comparisons. Each row belongs either to the genetic or activity type comparison, as indicated by the comparison type column. The ``Category 1" and ``Category 2" columns indicate the categories being compared, where each displays the mean and standard deviation of a given breath statistic either pre- or post-sigh. The difference between categorical distributions is evaluated using the p-value and difference between the categorical means. The sigh impact is the difference between the pre- and post-sigh p-values.}
    \label{table:sigh_pleth_res}
\end{table*}
    
\begin{table*}[h!]
    \centering
    \includegraphics[width=10cm]{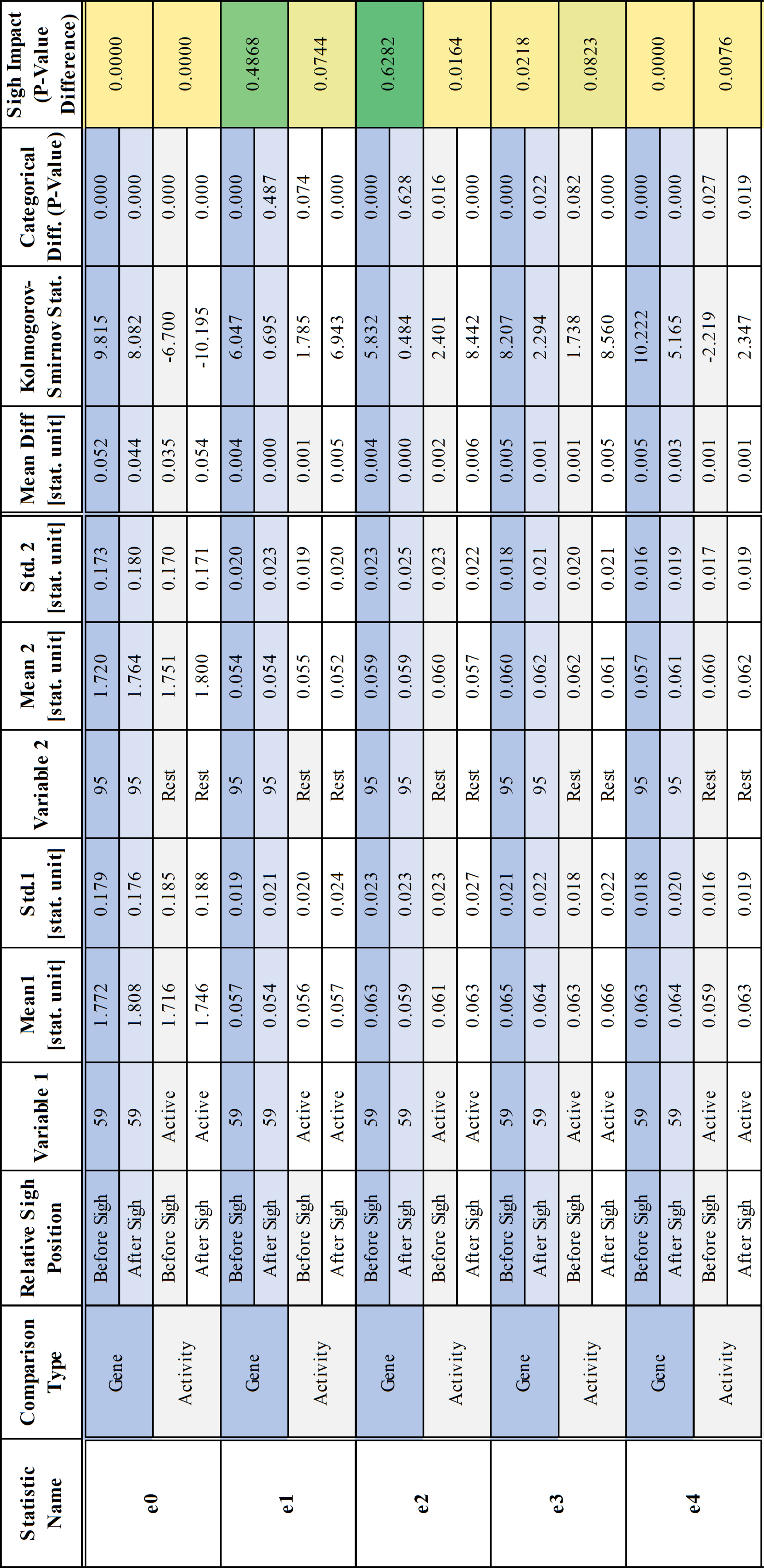}
    \caption{Comparison of the entropy statistics before and after the rat sigh for both genetic and activity type comparisons. Each row belongs either to the genetic or activity type comparison, as indicated by the comparison type column. The ``Category 1" and ``Category 2" columns indicate the categories being compared, where each displays the mean and standard deviation of a given entropy statistic either pre- or post-sigh. The difference between categorical distributions is evaluated using the p-value and difference between the categorical means. The sigh impact is the difference between the pre- and post-sigh p-values}
    \label{table:sigh_ent_res}
\end{table*}

\bibliographystyle{plain}
\bibliography{references}

\end{document}